# THE TWO-BODY INTERACTION WITH A CIRCLE IN TIME

Sarah B. M. Bell,[1,2] John P. Cullerne,[1] Bernard M. Diaz[1,3]

## Abstract

We complete our previous[1,2] demonstration that there is a family of new solutions to the photon and Dirac equations using spatial and temporal circles and four-vector behaviour of the Dirac bispinor. We analyse one solution for a bound state, which is equivalent to the attractive two-body interaction between a charged point particle and a second, which remains at rest. We show this yields energy and angular momentum eigenvalues that are identical to those found by the usual method of solving of the Dirac equation,[5] including fine structure. We complete our previous derivation[3] of QED from a set of rules for the two-body interaction and generalise these. We show that QED may be decomposed into a two-body interaction at every point in spacetime.

[1] Department of Computer Science I.Q. Group, The University of Liverpool, Chadwick Building, Peach Street, Liverpool, L69 7ZF, United Kingdom.

[2] U.K. phone number and email address, 01865 798579 and Sarabell@dial.pipex.com

[3] E-mail, B.M.Diaz@csc.liv.ac.uk

## 1. INTRODUCTION

### 1.1 Note on nomenclature

$\ddagger$ signifies quaternion conjugation. A lowercase Latin subscript stands for 1, 2 or 3 and indicates the space co-ordinates. A lowercase Greek subscript stands for 0, 1, 2, or 3 and indicates the spacetime co-ordinates. $i = \sqrt{-1}$. $\mathbf{i}_0 = 1$. $\mathbf{i}_1 = \mathbf{i}, \mathbf{i}_2 = \mathbf{j}, \mathbf{i}_3 = \mathbf{k}$ stand for the quaternion matrices, where $\mathbf{i}_r^2 = -1$, $\mathbf{i}_1\mathbf{i}_2 = \mathbf{i}_3, \mathbf{i}_2\mathbf{i}_1 = -\mathbf{i}_3$ with cyclic variations. We have $\mathbf{i}_0^\ddagger = \mathbf{i}_0, \mathbf{i}_r^\ddagger = -\mathbf{i}_r$. We set the speed of light, $c$, and $h/2\pi$, where $h$ is Planck's constant, to 1 except for final results.

### 1.2 Motivation

We showed that there is a family of new solutions to the photon and Dirac equations by mapping them onto a new space, consisting of a spatial circle.[2] Here we add another new space, consisting of a temporal circle, which leads to further new solutions.

One of the solutions, which we call *the coupled interaction*, is equivalent to a bound state formed by a charged point body obeying the Dirac equation. The body moves in an inverse-distance potential described by the photon equation and produced by another oppositely charged point body, which remains at rest. We call this *the two-body interaction*. This physical problem has been well studied and the possible eigenvalues, found by solving the Dirac equation in the usual way, are known with certainty.[5] We call this model *the Dirac interaction*. The coupled interaction described here is therefore either unphysical, which would be unwelcome, or must yield the known values. An analysis shows that it conforms under a consistent set of assumptions. However, it has new symmetries, discussed elsewhere[2] and generalised here, which could lead to new physics.

We also discussed a new method of quantising the classical electromagnetic field by quantising the two-body interaction,[3] which could lead to new applications. We expand this derivation of QED by including the full spectrum of possible eigenstates as a basis.

## 2. DIRAC AND PHOTON EQUATIONS ON NEW SPACES

### 2.1 Preliminaries

We introduce the versatile form of the Dirac equation,[1] which we use throughout. There is also a versatile form of the photon equation with the same transformational behaviour.[1]

The versatile form of the Dirac equation for an electron is

$$(\underline{\mathbf{D}} - ie\underline{\mathbf{A}}^\sim)\underline{\boldsymbol{\Phi}} = \underline{\boldsymbol{\Phi}}\underline{\mathbf{M}}^\sim \qquad (2.1.\text{A})$$

where $e$ is the electric charge on the electron, and we define the meaning of the underline by

$$\underline{\mathbf{D}} = \underline{\mathbf{D}}(\mathbf{D}, \mathbf{D}^\ddagger) = \begin{pmatrix} 0 & \mathbf{D} \\ \mathbf{D}^\ddagger & 0 \end{pmatrix} \qquad (2.1.\text{B})$$

We call such matrices *reflectors*. We define **D** by saying

$$\mathbf{D}_0 = i\mathbf{i}_0\, \partial/\partial x_0, \qquad \underline{\mathbf{D}}_0 = \underline{\mathbf{D}}_0(\mathbf{D}_0, \mathbf{D}_0^\ddagger) \qquad (2.1.\text{C})$$

$$\mathbf{D}_r = \mathbf{i}_r\, \partial/\partial x_r, \qquad \underline{\mathbf{D}}_r = \underline{\mathbf{D}}_r(\mathbf{D}_r, \mathbf{D}_r^\ddagger)$$

where the $x_\mu$ are real, and then

$$\mathbf{D} = \mathbf{D}_0 + \mathbf{D}_1 + \mathbf{D}_2 + \mathbf{D}_3, \qquad \underline{\mathbf{D}} = \underline{\mathbf{D}}_0 + \underline{\mathbf{D}}_1 + \underline{\mathbf{D}}_2 + \underline{\mathbf{D}}_3 \qquad (2.1.\text{D})$$

The Dirac equation, (2.1.A), and also versatile form of the photon equation hold on in the usual Minkowski spacetime, which we call *L*. We define $\underline{\mathbf{A}}^\sim$, the potential term, in a similar way to $\underline{\mathbf{D}}$

$$\underline{\mathbf{A}}^\sim = \underline{\mathbf{A}}^\sim(\mathbf{A}^\sim, \mathbf{A}^{\sim\ddagger}), \qquad \mathbf{A}^\sim = A_0/i + \mathbf{i}_1 A_1 + \mathbf{i}_2 A_2 + \mathbf{i}_3 A_3 \qquad (2.1.\text{E})$$

where the four-vector $(A_0, A_1, A_2, A_3)$ represents the potential. The wave function, $\underline{\Phi}$, is defined by

$$\underline{\Phi} = \underline{\Phi}(\phi_1, \phi_2) \qquad (2.1.\text{F})$$

where $\phi_1$ and $\phi_2$ are quaternions whose relation to the bispinors in the original version of the Dirac equation was described in our earlier paper.[1] $\underline{\mathbf{M}}^\sim$ is defined by

$$\underline{\mathbf{M}}^\sim = \underline{\mathbf{M}}^\sim(\mathbf{M}^\sim, -\mathbf{M}^{\sim\ddagger}) \qquad (2.1.\text{G})$$
$$\mathbf{M}^\sim = \mathbf{i}_0 M_0^\sim + \mathbf{i}_1 M_1 + \mathbf{i}_2 M_2 + \mathbf{i}_3 M_3, \qquad \mathbf{M}^\sim \mathbf{M}^{\sim\ddagger} = -(m^b)^2$$

where $m^b$ is the rest mass of the electron. As we explained,[1] for the usual half-angular behaviour of the phase of the bispinor, $\phi_1$, under rotation and the analogue under Lorentz transformation, $\mathbf{M}^\sim = -im^b$, a constant. $\phi_2$ also shows half-angular behaviour, but it rotates in the opposite sense to $\phi_1$ for the temporal rotations, which implement Lorentz transformations for the versatile Dirac equation. This means that $\phi_2^\ddagger$ rotates in the same sense as $\phi_1$. In this case the versatile Dirac equation is simply another form of the usual version. However, the versatile Dirac and photon equations also allow four-vector behaviour of the bispinors $\phi_1$ and $\phi_2^{\ddagger}$. In this case $(iM_0^{\sim}, M_1, M_2, M_3)$ behaves like the energy-momentum four-vector of the electron. It is this form of behaviour that we use here.

## 2.2. Dirac and photon equations on $T$

In our previous paper[2] we defined a space $M$ that could be mapped to $L$ and described how to obtain $M$ from $L$ by introducing a transformation called *the Circular transformation*, which we applied to $x_1$ and $x_2$. We call the plane addressed by these co-ordinates *the spatial plane*. We showed that if the Dirac and photon equations hold on $L$ they also hold on $M$. Here we apply the Circular transformation to $x_0$ and $x_3$, which results in a space we call $T$ that also maps to $L$. We call the plane addressed by $x_0$ and $x_3$ *the temporal plane*.

Except for the light cone, the temporal plane of $L$ may be described using co-ordinates $(r_0, \theta_0)$, $r_0$ and $\theta_0$ real, where

$$x_0 = r_0 \sinh\theta_0, \qquad x_3 = r_0 \cosh\theta_0 \qquad (2.2.A)$$

$$\overrightarrow{\delta \tilde{s}_0} = r_0 \overrightarrow{\delta \theta_0}, \qquad \tilde{s}_0 = r_0 \theta_0 \qquad (2.2.B)$$

where $\tilde{s}_0$ is real. Using the standard method

$$\begin{pmatrix} \dfrac{\partial}{\partial x_0} \\ \dfrac{\partial}{\partial x_3} \end{pmatrix} = \begin{pmatrix} \cosh\theta_0 & -\sinh\theta_0 \\ -\sinh\theta_0 & \cosh\theta_0 \end{pmatrix} \begin{pmatrix} \dfrac{1}{r_0}\dfrac{\partial}{\partial \theta_0} \\ \dfrac{\partial}{\partial r_0} \end{pmatrix} \qquad (2.2.C)$$

This has the form of a Lorentz transformation. As we discussed earlier,[1]

we may introduce an imaginary angle, $\hat{\theta}_0 = -i\theta$, and obtain

$$\begin{pmatrix} i\dfrac{\partial}{\partial x_0} \\ \dfrac{\partial}{\partial x_3} \end{pmatrix} = \begin{pmatrix} \cos\hat{\theta}_0 & \sin\hat{\theta}_0 \\ -\sin\hat{\theta}_0 & \cos\hat{\theta}_0 \end{pmatrix} \begin{pmatrix} \dfrac{i}{r_0}\dfrac{\partial}{\partial \theta_0} \\ \dfrac{\partial}{\partial r_0} \end{pmatrix} \quad (2.2.D)$$

We see that $-i\tilde{s}_0$ is the arc corresponding to $r_0$ and $\hat{\theta}_0$ in $L$ from the second of equations (2.2.B). We introduce a further vector along the arc

$$\overrightarrow{\delta s_0} = R_0 \overrightarrow{\delta \theta_0}, \qquad s_0 = R_0 \theta_0 \quad (2.2.E)$$

where $R_0$ is constant. Using the second of equations (2.2.E) and (2.2.B), equation (2.2.D) becomes

$$i\dfrac{\partial}{\partial x_0} = \sin\hat{\theta}_0 \dfrac{\partial}{\partial r_0} + i\dfrac{s_0}{\tilde{s}_0}\cos\hat{\theta}_0 \dfrac{\partial}{\partial s_0}, \quad (2.2.F)$$

$$\dfrac{\partial}{\partial x_3} = \cos\hat{\theta}_0 \dfrac{\partial}{\partial r_0} - i\dfrac{s_0}{\tilde{s}_0}\sin\hat{\theta}_0 \dfrac{\partial}{\partial s_0}$$

Again excepting the light cone, we transform the four-vectors associated with **A** and **M** with

$$-iA_0 = A_{r_0}\sin\hat{\theta}_0 + A_{\tilde{s}_0}\cos\hat{\theta}_0, \quad (2.2.G)$$

$$A_3 = A_{r_0}\cos\hat{\theta}_0 - A_{\tilde{s}_0}\sin\hat{\theta}_0,$$

$$M_{\tilde{0}} = M_{r_0}\sin\hat{\theta}_0 + M_{\tilde{s}_0}\cos\hat{\theta}_0,$$

$$M_3 = M_{r_0}\cos\hat{\theta}_0 - M_{\tilde{s}_0}\sin\hat{\theta}_0$$

where $A_{r_0}$ and $M_{r_0}$ are the components along the radius, $\overrightarrow{\delta r_0}$, and $A_{\tilde{s}_0}$ and $M_{\tilde{s}_0}$ are the components along the arc, $\overrightarrow{-i\delta s_0}$. We transform the reflector matrices, $\underline{\mathbf{i}}_0\left(\mathbf{i}_0, \mathbf{i}_0^{\ddagger}\right)$ and $\underline{\mathbf{i}}_3\left(\mathbf{i}_3, \mathbf{i}_3^{\ddagger}\right)$, with

*The two-body interaction with a circle in time*

$$\underline{\mathbf{i}}_0 = \underline{\mathbf{i}}_{r_0} \sin\hat{\theta}_0 + \underline{\mathbf{i}}_{s_0} \cos\hat{\theta}_0, \quad (2.2.\text{H})$$
$$\underline{\mathbf{i}}_3 = \underline{\mathbf{i}}_{r_0} \cos\hat{\theta}_0 - \underline{\mathbf{i}}_{s_0} \sin\hat{\theta}_0$$

where $\underline{\mathbf{i}}_{r_0}$ and $\underline{\mathbf{i}}_{s_0}$ are the reflector matrices along the radius, $\overrightarrow{\delta r_0}$, and the arc, $\overrightarrow{-i\delta s_0}$, respectively. The anti-commutation relations of the matrices $\underline{\mathbf{i}}_{s_0}, \underline{\mathbf{i}}_1, \underline{\mathbf{i}}_2$ and $\underline{\mathbf{i}}_{r_0}$ are the same as that of the matrices $\underline{\mathbf{i}}_0, \underline{\mathbf{i}}_1, \underline{\mathbf{i}}_2$ and $\underline{\mathbf{i}}_3$, respectively,

$$\underline{\mathbf{i}}_\mu^2 = 1, \qquad \underline{\mathbf{i}}_\mu \underline{\mathbf{i}}_\nu = -\underline{\mathbf{i}}_\nu \underline{\mathbf{i}}_\mu \text{ where } \underline{\mathbf{i}}_\mu \neq \underline{\mathbf{i}}_\nu \quad (2.2.\text{I})$$

There is no distinction between spatial and temporal co-ordinates in these anti-commutation relations leaving the multiplication of the temporal co-ordinates by $i$ as the only remaining difference. From equations (2.1.E), (2.1.G), (2.2.G) and (2.2.H) we obtain

$$\underline{\mathbf{A}}^{\sim} = \underline{\mathbf{i}}_{s_0} A_{s_0}^{\sim} + \underline{\mathbf{i}}_1 A_1 + \underline{\mathbf{i}}_2 A_2 + \underline{\mathbf{i}}_{r_0} A_{r_0}, \quad (2.2.\text{J})$$
$$\underline{\mathbf{M}}^{\sim} = \underline{\mathbf{i}}_{s_0} M_{s_0}^{\sim} + \underline{\mathbf{i}}_1 M_1 + \underline{\mathbf{i}}_2 M_2 + \underline{\mathbf{i}}_{r_0} M_{r_0}$$

When $r_0 = R_0$ and $\tilde{s}_0 = s_0$ equations (2.1.C), (2.2.F) and (2.2.H) give

$$\underline{\mathbf{D}}_0 + \underline{\mathbf{D}}_3 = i\underline{\mathbf{i}}_{s_0} \partial/\partial s_0 + \underline{\mathbf{i}}_{r_0} \partial/\partial r_0 \quad (2.2.\text{K})$$

while $\underline{\mathbf{D}}_1$ and $\underline{\mathbf{D}}_2$ are unchanged, in a pattern similar to that of equations (2.2.J). We would like equation (2.2.K) to hold universally, as would be the case if equations (2.2.F) were altered so that the first term on the right-hand-side still applied but the multiplier, $s_0/\tilde{s}_0$, of the second term was removed. $\overrightarrow{\delta r_0}$ is orthogonal to both $\overrightarrow{\delta \tilde{s}_0}$ and $\overrightarrow{\delta s_0}$, from the first of equations (2.2.B) and (2.2.E), a feature we wish to perpetuate. We may obtain the effect we require by defining a new space, *T*, in which we retain $\vec{\delta r_0}$ but set $\vec{\delta \tilde{s}_0}$ to $\vec{\delta s_0}$. The second of equations (2.2.B) and (2.2.E) then provide a bijection between *L*, co-ordinated by $(\tilde{s}_0, x_1, x_2, r_0)$, and *T*, co-ordinated by $(s_0, x_1, x_2, r_0)$, since

$$\tilde{s}_0 = \frac{r_0 s_0}{R_0} \tag{2.2.L}$$

which we extend to include the light cone. From equations (2.2.J) and (2.2.K) we can translate the Dirac equation into *T* immediately. It has the same form as equation (2.1.A).

It is easy to translate the photon equation into the variables appropriate for *T*, but we need to consider whether the volume element used in the equation is the same in spaces *L* and *T* for the reasons we discussed previously.[2] The volume element for *L* is

$$\delta V^L = \delta x_1 \delta x_2 \delta x_3 \tag{2.2.M}$$

We consider the volume element where $\vec{\delta x_3} = \vec{\delta r_0}$, obtaining

$$\delta V^L = \delta x_1 \delta x_2 \delta r_0 \tag{2.2.N}$$

This is the same as the volume element for *T*. Therefore any potential is the same for both spaces, since the volume elements are constant in each.

We note that the Dirac and photon equations have the same form and significance for *T* as for *L* after the map

$$s_0 \to x_0, \qquad r_0 \to x_3 \tag{2.2.O}$$

Since $s_0$ describes a circle, we see that $T$ differs topologically from $L$. Apart from this, the co-ordinates $(s_0, x_1, x_2, r_0)$ appear to an observer in $T$ to be a Cartesian system in a flat spacetime. This completes the description of the Circular transformation for the temporal plane.

### 2.3 Dirac and photon equations on $S$

We state our previous results for $M^{(2)}$ in the terminology we use here. The spatial plane of $L$ may be described using polar co-ordinates $(r_1, \theta_1)$, where

$$x_1 = r_1 \sin\theta_1, \qquad x_2 = r_1 \cos\theta_1 \qquad (2.3.\text{A})$$

$$\overrightarrow{\delta\tilde{s}_1} = r_1 \overrightarrow{\delta\theta_1}, \qquad \tilde{s}_1 = r_1 \theta_1 \qquad (2.3.\text{B})$$

and $\tilde{s}_1$ is the arc corresponding to $r_1$ and $\theta_1$ in $L$. We introduce a further vector along the arc

$$\overrightarrow{\delta s_1} = R_1 \overrightarrow{\delta\theta_1}, \qquad s_1 = R_1 \theta_1 \qquad (2.3.\text{C})$$

where $R_1$ is constant.

We transform the four-vectors associated with $\tilde{\mathbf{A}}$ and $\tilde{\mathbf{M}}$ with

$$A_1 = A_{r_1} \sin\theta_1 + A_{s_1} \cos\theta_1, \quad A_2 = A_{r_1} \cos\theta_1 - A_{s_1} \sin\theta_1 \qquad (2.3.\text{D})$$
$$M_1 = M_{r_1} \sin\theta_1 + M_{s_1} \cos\theta_1, \quad M_2 = M_{r_1} \cos\theta_1 - M_{s_1} \sin\theta_1$$

where $A_{r_1}$ and $M_{r_1}$ are the components along the radius, $\overrightarrow{\delta r_1}$, and $A_{s_1}$ and $M_{s_1}$ are the components along the arc, $\overrightarrow{\delta s_1}$. We transform the reflector matrices, $\underline{\mathbf{i}}_1(\mathbf{i}_1, \mathbf{i}_1^{\ddagger})$ and $\underline{\mathbf{i}}_2(\mathbf{i}_2, \mathbf{i}_2^{\ddagger})$, with

$$\underline{\mathbf{i}}_1 = \underline{\mathbf{i}}_{r_1} \sin\theta_1 + \underline{\mathbf{i}}_{s_1} \cos\theta_1, \qquad (2.3.\text{E})$$
$$\underline{\mathbf{i}}_2 = \underline{\mathbf{i}}_{r_1} \cos\theta_1 - \underline{\mathbf{i}}_{s_1} \sin\theta_1$$

where $\underline{\mathbf{i}}_{r_1}$ and $\underline{\mathbf{i}}_{s_1}$ are the reflector matrices along the radius, $\overrightarrow{\delta r_1}$, and the arc, $\overrightarrow{\delta s_1}$, respectively. The anti-commutation relations of the matrices $\underline{\mathbf{i}}_0, \underline{\mathbf{i}}_{s_1}, \underline{\mathbf{i}}_{r_1}$ and $\underline{\mathbf{i}}_3$ are the same as those of the matrices $\underline{\mathbf{i}}_0, \underline{\mathbf{i}}_1, \underline{\mathbf{i}}_2$ and $\underline{\mathbf{i}}_3$, respectively. We have

$$\underline{\mathbf{A}}\tilde{} = -i\underline{\mathbf{i}}_0 A_0 + \underline{\mathbf{i}}_{s_1} A_{s_1} + \underline{\mathbf{i}}_{r_1} A_{r_1} + \underline{\mathbf{i}}_3 A_3, \qquad (2.3.\text{F})$$
$$\underline{\mathbf{M}}\tilde{} = \underline{\mathbf{i}}_0 M_0\tilde{} + \underline{\mathbf{i}}_{s_1} M_{s_1} + \underline{\mathbf{i}}_{r_1} M_{r_1} + \underline{\mathbf{i}}_3 M_3$$

We define the new space, $M$, by retaining $\overrightarrow{\delta r_1}$ but setting $\overrightarrow{\delta \tilde{s}_1}$ to $\overrightarrow{\delta s_1}$. The second of equations (2.3.B) and (2.3.C) then provide a bijection between $L$, co-ordinated by $(x_0, \tilde{s}_1, r_1, x_3)$, and $M$, co-ordinated by $(x_0, s_1, r_1, x_3)$, since

$$\tilde{s}_1 = \frac{r_1 s_1}{R_1} \qquad (2.3.\text{G})$$

In $M$

$$\underline{\mathbf{D}}_1 + \underline{\mathbf{D}}_2 = \underline{\mathbf{i}}_{s_1} \partial/\partial s_1 + \underline{\mathbf{i}}_{r_1} \partial/\partial r_1 \qquad (2.3.\text{H})$$

while $\underline{\mathbf{D}}_0$ and $\underline{\mathbf{D}}_3$ are unchanged. From equations (2.3.F) and (2.3.H) we can translate the Dirac equation into $M$ immediately.

It is easy to translate the photon equation into the variables appropriate for $M$. However, for $M$ the volume element is different, so that we have for the potential term in $M$

$$\underline{\mathbf{A}}^{B\sim} = \frac{r_1 \underline{\mathbf{A}}^{\sim}}{R_1} \tag{2.3.I}$$

We note that the Dirac and photon equations have the same form and significance for *M* as for *L* after the map

$$s_1 \to x_1, \qquad r_1 \to x_2, \qquad \underline{\mathbf{A}}^{B\sim} \to \underline{\mathbf{A}}^{\sim} \tag{2.3.J}$$

*M* still differs topologically from *L*. Apart from this, the co-ordinates $(x_0, s_1, r_1, x_3)$ appear to an observer in *M* to be a Cartesian system in a flat spacetime. This completes the description of the Circular transformation for the spatial plane.

The temporal and spatial planes constitute two independent subspaces of *L* and so we may apply the Circular transformation to both planes and then join the result to obtain a four-dimensional space of a new type we call *S*. *L* is co-ordinated by $(\tilde{s}_0, \tilde{s}_1, r_1, r_0)$, while *S* is co-ordinated by $(s_0, s_1, r_1, r_0)$, and they are related by the bijection given in equations (2.2.L) and (2.3.G). For *S*

$$\underline{\mathbf{A}}^{\sim} = \underline{\mathbf{i}}_{s_0} A_{s_0}^{\sim} + \underline{\mathbf{i}}_{s_1} A_{s_1} + \underline{\mathbf{i}}_{r_1} A_{r_1} + \underline{\mathbf{i}}_{r_0} A_{r_0}, \tag{2.3.K}$$
$$\underline{\mathbf{A}}^{B\sim} = \underline{\mathbf{i}}_{s_0} A_{s_0}^{B\sim} + \underline{\mathbf{i}}_{s_1} A_{s_1}^{B} + \underline{\mathbf{i}}_{r_1} A_{r_1}^{B} + \underline{\mathbf{i}}_{r_0} A_{r_0}^{B},$$
$$\underline{\mathbf{M}}^{\sim} = \underline{\mathbf{i}}_{s_0} M_{s_0}^{\sim} + \underline{\mathbf{i}}_{s_1} M_{s_1} + \underline{\mathbf{i}}_{r_1} M_{r_1} + \underline{\mathbf{i}}_{r_0} M_{r_0}$$

and

$$\underline{\mathbf{D}}_0 + \underline{\mathbf{D}}_1 + \underline{\mathbf{D}}_2 + \underline{\mathbf{D}}_3 = \tag{2.3.L}$$
$$i\underline{\mathbf{i}}_{s_0} \partial/\partial s_0 + \underline{\mathbf{i}}_{s_1} \partial/\partial s_1 + \underline{\mathbf{i}}_{r_1} \partial/\partial r_1 + \underline{\mathbf{i}}_{r_0} \partial/\partial r_0$$

The final version of the Dirac equation for *S* is then

$$\{\underline{\mathbf{D}}(\mathbf{D},\mathbf{D}^{\ddagger})-ie\underline{\mathbf{A}}^{B\sim}(\mathbf{A}^{B\sim},\mathbf{A}^{B\sim\ddagger})\}\underline{\Phi}^{B}(\phi_{1}^{B},\phi_{2}^{B}) \qquad (2.3.\text{M})$$
$$=\underline{\Phi}^{B}(\phi_{1}^{B},\phi_{2}^{B})\underline{\mathbf{M}}^{\sim}(\mathbf{M}^{\sim},-\mathbf{M}^{\sim\ddagger})$$

where $\underline{\Phi}^{B}$ is the solution associated with the new potential term $\underline{\mathbf{A}}^{B\sim}$. The Dirac and photon equations have the same form and significance for *S* as for *L* after the map

$$s_0 \to x_0, \qquad s_1 \to x_1, \qquad r_1 \to x_2, \qquad (2.3.\text{N})$$
$$r_0 \to x_3, \qquad \mathbf{A}^{B\sim} \to \mathbf{A}^{\sim}$$

*S* differs topologically from *L*. Apart from this the co-ordinates $(s_0, s_1, r_1, r_0)$ appear to an observer in *S* to be Cartesian co-ordinates applied to a flat spacetime. This means that frames in *S*, which become Cartesian under the mapping (2.3.N), are inertial even though they are attached to bodies rotating round the origin in *L*. This also applies to *T* and *M* under the mappings (2.2.O) and (2.3.J), respectively.

We add a superscript *y* to our co-ordinates, obtaining for $S(s_0^y, s_1^y, r_1^y, r_0^y)$, where *y* indicates the particle referred to. We also need the variant

$$s_0^{y\sim} = s_0^y/i, \qquad s_1^{y\sim} = s_1^y/i \qquad (2.3.\text{O})$$

## 3.   SOLVING THE PHOTON AND DIRAC EQUATIONS

### 3.1   Temporal circle

We saw in section 2.2 that the formation of the circular temporal orbit described by the Dirac equation in *T* did not necessitate any change to the potential that held in *L*. We suppose that a stable temporal circle forms in the path of a free fermion, which we will take to be an electron. We call this *the circle electron*. We will describe the straight path of the circle electron in the rest frame in *L* by the temporal co-ordinate $x_0$ and the circular path in *T* by $s_0^l$. $x_0$ is tangent to $s_0^l$ at a point *P*. The electron enters the circle at *P* and may circulate the circle one or more times, eventually exiting, also at *P*. For this process to be invisible to observers in *L*, the energy of the electron in *L* may not change and the phase of the electron wave function at *P* must be single-valued. Further, an observer in *L* must not be able to detect the circle electron in *T*.

The above is only true for the first-quantised description we apply throughout here, since, for example, the finite length of the circle will also induce the equivalent of a Casmir effect in the vacuum.[9] However, it appears arbitrary to assume that an observer in *L* cannot detect the circle electron in *T*, given the bijection connecting the two spaces in equation (2.2.L). We therefore digress to discuss appearances should we assume the opposite. For a given instant the electron appears in general at a location, $\hat{P}$, in *L*, which we shall address as $(\hat{x}_0, 0, 0, 0)$, and possibly in two locations, $P^+$ and $P^-$, on the circle in *T*. We may address these locations in *L* as $(\hat{x}_0, x_1, 0, 0)$ and $(\hat{x}_0, -x_1, 0, 0)$. However, the electron is travelling in opposite temporal directions at $P^+$ and $P^-$ for the observer in *L* and will therefore appear as an electron-positron pair[5] created close to the single electron at $\hat{P}$. Unless sufficient energy is present the electron-positron pair will be virtual, one of those subsumed in the renormalisation series for the electron at $\hat{P}$ and therefore invisible if we assume the physical values for the charge and mass.[5]

To first order, we find the implications of the temporal circle by solving the free Dirac equation (2.1.A) for $T$, obtaining

$$\phi_1^l = \exp(im^{b\sim} s_0^{l\sim}), \qquad \phi_2^l = m^{b\sim}(\mathbf{M}^\sim)^{-1} \exp(im^{b\sim} s_0^{l\sim}) \qquad (3.1.\text{A})$$

where $m^{b\sim} = m^b/i$, equations (2.3.O) hold with superscript $y = l$, and we have a plane wave solution, $\underline{\Phi} = \underline{\Phi}^l(\phi_1^l, \phi_2^l)$, with $im^{b\sim}$ the frequency. Let the radius of the temporal circle, the parameter $R_0$ in the mapping given in equation (2.2.L), be $R_0^l$. If we set $\theta_0 = 2\pi$ and $s_0 = s_0^l$, the second of equations (2.2.E) provides the length, $2\pi R_0^l$, of the path $s_0^l$ traces round the origin in $T$. Then the condition the phase is single-valued at $P$ is

$$m^{b\sim} R_0^{l\sim} = n_\theta \frac{h}{2\pi} \qquad (3.1.\text{B})$$

where $-iR_0^{l\sim} = R_0^l$, we have restored $h/2\pi$ and taken $n_\theta$ as a positive integer. Equation (3.1.B) quantises the length of the circle.

We suggest the circle itself may also vibrate if sufficient energy is present, in parallel with, but without directly employing, string theory.[9] We will call these standing waves *circle waves* and the interaction supplying the energy *the circle interaction*. We treat the circle waves as neutral quasi-particles obeying the free Dirac equation, (2.1.A), in $T$ since, first, the circle waves do not interact with the circle electron but, second, will be incorporated into the final coupled-interaction solution of the equation for another electron, called *the coupling electron*. By a derivation similar to the one used for the circle electron, the energy, $i\eta^{l\sim}$, associated with a circle wave, is

$$\eta^{l\sim} = \frac{n_r h}{2\pi R_0^{l\sim}} \qquad (3.1.C)$$

where $n_r$ is a positive integer. We will consider only one, since summing with the energies of further waves adds nothing new. Equations (3.1.B) and (3.1.C) allow us to obtain an expression for the energy contributed by the circle wave

$$\eta^{l\sim} = \frac{n_r m^{b\sim}}{n_\theta} \qquad (3.1.D)$$

The energy required to form a circle wave can be supplied by a potential due to a second charge in *L*, but the inclusion of a second charge also generates a spatial circle,[2] which changes the space from *T* to a space of type *S*.

### 3.2 Spatial circle

We call the interaction associated with the spatial circle *the Bohr interaction*. Our description consists of summarising the solution of equation (2.3.M) for an electron with co-ordinates $(s_0^b, s_1^b, r_1^b, r_0^b)$, in a space of type *S* we shall call $S^b$. The co-ordinates have the same significance as $(s_0, s_1, r_1, r_0)$, the superscript *b* indicating the particle to which the co-ordinates apply. The solution is identical to the one in *M* we discussed previously[2] except that here we have $s_0^b$ in place of $x_0$ and $r_0^b$ in place of $x_3$. The solution may be written

*The two-body interaction with a circle in time*

$$\phi_1^b = \exp\{i(v^{b\sim} s_0^{b\sim} + \mu^b s_1^b)\}, \qquad (3.2.A)$$

$$\phi_2^b = \{iv^{b\sim} - \ddot{\mathbf{i}}_{s_1}\mu^b - ieA^{b\sim}\}(\mathbf{M}^\sim)^{-1} \exp\{i(v^{b\sim} s_0^{b\sim} + \mu^b s_1^b)\},$$

$$(m^{b\sim})^2 = (v^{b\sim} - eA^{b\sim})^2 + (\mu^b)^2$$

where $m^{b\sim} = m^b/i$, equations (2.3.O) hold with superscript $y = b$, and we have a plane wave solution $\underline{\Phi}^B = \underline{\Phi}^b(\phi_1^b, \phi_2^b)$, where $iv^{b\sim}$ is the frequency and $\mu^b$ the wave number. $iA^{b\sim}$ is the potential due to the particle attracting the electron, which we call *the nucleus*, in the rest frame of the nucleus. We define the potential so that it corresponds to an inverse-distance potential in *L*, using equation (2.3.I),

$$A^{b\sim} = ie/R_1^b \qquad (3.2.B)$$

where *e* is the charge on the nucleus and we have set the parameter $R_1$, in the mapping given in equation (2.3.G), to $R_1^b$. We call $R_1^b$ *the Bohr radius*.

The third of equations (3.2.A) defines the equivalent of a free electron with wave number $\mu^b$ and frequency $i\eta^{b\sim}$ where

$$\eta^{b\sim} = v^{b\sim} - eA^{b\sim} \qquad (3.2.C)$$

We identify this as *the Bohr electron*. We suppose that the presence of the potential required for the spatial orbit has imparted a velocity, $v^b$, to the circle electron, which then becomes the Bohr electron just described. The temporal circle becomes part of a bound state of the electron at this point and, while the rest frames of the circle wave and nucleus are the same, the rest frame of the electron has changed.

We may solve the Dirac equation (2.3.M) with zero potential for the Bohr electron, which validates de Broglie's relations.[8] The first relation is

$$\eta^{b\sim} = \frac{m^{b\sim}}{\sqrt{1+(v^{b\sim})^2}} \quad (3.2.D)$$

where $-iv^{b\sim} = v^b$, while the second is

$$\mu^b = \frac{m^{b\sim} v^{b\sim}}{\sqrt{1+(v^{b\sim})^2}} \quad (3.2.E)$$

We find the energy of the Bohr interaction four-vector, which has the same rest frame as the nucleus, rather than that of the electron. In the rest frame of the electron the energy is simply $im^{b\sim}$, since there is no potential, and so

$$m^{b\sim} = \frac{v^{b\sim}}{\sqrt{1+(v^{b\sim})^2}} \quad (3.2.F)$$

Equations (3.2.B), (3.2.C), (3.2.D) and (3.2.F) give

$$\frac{ie^2}{R_1^b} = \frac{m^{b\sim}(v^{b\sim})^2}{\sqrt{1+(v^{b\sim})^2}} \quad (3.2.G)$$

This circular motion resembles the orbits of Bohr[6, 7] and we call the equation *Bohr's first equation*. It is the same as Bohr's but, unlike Bohr, we transfer it to the new space $S^b$.

### 3.3 Combination of a spatial and temporal circle

We require the relativistically invariant expression of equation (3.1.B) to describe the circle as it now applies for the Bohr electron with the new rest frame. If we set $\theta_0 = \theta_1 = 2\pi$, the second of equations (2.2.E) and (2.3.C) provide the lengths of the paths $s_0$ and $s_1$ trace round the origin in $S$. We will denote these lengths in $S^b$ for the Bohr electron by $-2i\pi R_0^{b\sim}$ and $2\pi \hat{R}_1^b$. After the mapping in (2.3.N), $s_0^b$ and $s_1^b$ behave like the first two elements of a four-vector and therefore so do $-2i\pi R_0^{b\sim}$ and $2\pi \hat{R}_1^b$. This means that the relativistically invariant form of equation (3.1.B) is

$$\eta^{b\sim} R_0^{b\sim} + \mu^b \hat{R}_1^b = n_\theta \frac{h}{2\pi}, \qquad (3.3.A)$$

$$R_0^{b\sim} = \frac{R_0^{l\sim}}{\sqrt{1+(v^{b\sim})^2}}, \qquad \hat{R}_1^b = \frac{v^{b\sim} R_0^{l\sim}}{\sqrt{1+(v^{b\sim})^2}}$$

where the result holds because the left-hand-side of equation (3.1.B) and the first of equations (3.3.A) both represent the dot product of the same four-vectors. Relativistic invariance requires more, that the electron spin should behave like a four-vector, as we have already discussed for a circular spatial orbit.[2] This behaviour can be explained by taking into account the alteration of phase due to the circular motion present.[2, 5] These remarks on spin also apply to the quasi-particle associated with the circle wave.

The spatial part of the circle has so far not been quantised individually, but for conformity with the usual solution of the Dirac or Schröedinger equation for a bound state,[5, 8] we must posit that the probability density be single-valued at every spatial location. Since the bispinors, $\phi_1$ and $\phi_2^{\ddagger}$, behave like a four-vectors, the wave function, $\underline{\boldsymbol{\Phi}}^b$, must be single-valued. This is equivalent to demanding an angular momentum eigenstate[2] and a measurement of this constrains the wave function and alters the length of the spatial circle again. With this proviso we may quantise the spatial circle as we did previously[2] and derive Bohr's second equation from equations (3.2.A)

$$\frac{\tilde{m}\, v^{b\sim} R_1^b}{\sqrt{1+\left(v^{b\sim}\right)^2}} = n_\theta \frac{h}{2\pi} \quad (3.3.\text{B})$$

where both sides are equal to the angular momentum in $L$.[2] The equation shows that the spatial circle contains a standing wave and is the same as that derived by Bohr. We call it *Bohr's second equation*. Again, we have transferred it to the new space $S^b$. The Bohr equations, (3.2.G) and (3.3.B), are a complete solution since they lead to an expression for the velocity of the electron

$$\frac{v^{b\sim} c}{i} = \frac{2\pi e^2}{n_\theta h} = \frac{\alpha c}{n_\theta} \approx \frac{c}{137 n_\theta} \quad (3.3.\text{C})$$

where $\alpha$ is the fine structure constant[5, 8] and we have restored $c$ as well as $h/2\pi$.

We turn to the temporal circle for the Bohr electron, described in the first term on the left-hand-side of the first of equations (3.3.A). We sum the electromagnetic energy and the kinetic energy of the Bohr electron as we did in equation (3.2.C) and consider instead of $i\eta^{b\sim}$ the total energy of the electron, $iv^{b\sim}$. We obtain from equations (3.1.B), (3.2.F) and the second of equations (3.3.A)

$$v^{b\sim} R_0^{b\sim} = n_\theta \frac{h}{2\pi} \qquad (3.3.D)$$

The temporal circle for the Bohr electron also contains a standing wave and the final definition of the parameters $R_0$ and $R_1$ for the space $S^b$ is $-iR_0^{b\sim}$ and $R_1^b$. We have now completed our summary of the Bohr interaction.

Since the circle wave is neutral the potential responsible for the Bohr interaction has no effect and equation (3.1.C) remains valid. From the rest frame of the Bohr electron, the wave now contains momentum as well as energy, but from the rest frame of the nucleus it continues to contain energy alone. Although the temporal circle also contains a standing circle wave, it does not occupy the same space as the Bohr electron, since for the wave and $T$, $R_0 = -iR_0^{l\sim}$, while for the Bohr electron and $S^b$, $R_0 = -iR_0^{b\sim}$. Before we demonstrate that we can find a common space, with the energy of the circle wave supplied by an electromagnetic potential, we need to derive some further properties of the versatile Dirac and photon equations.

## 4. TACHYONIC TRANSFORMATION

### 4.1 Tachyon Dirac and photon equations

We demonstrate that, corresponding to every solution of the versatile Dirac equation, (2.1.A) or (2.3.M), and the versatile photon equation, there is a tachyon solution in which both the spatial co-ordinate, $s_1$, and the temporal co-ordinate, $s_0$, are multiplied by $-i$ and then exchanged. We do not encounter any difficulty with causality in the tachyon solutions we use since these tachyons are part of a bound state and not apparent externally without destroying the state.

Suppose we have any four-vector, $\mathbf{X} = (X_0, X_1, X_2, X_3)$, $X_\mu$ real, in $L$, addressed by Cartesian co-ordinates $(x_0, x_1, x_2, x_3)$ with $x_\mu$ real. We use the Lorentz transformations to find $\mathbf{X}$ in a dashed frame with relative velocity $-v$ along $x_1$,

$$X_0'^{\sim} = \frac{X_0 + vX_1}{\sqrt{1-v^2}}, \qquad X_1'^{\sim} = \frac{X_1 + vX_0}{\sqrt{1-v^2}} \qquad (4.1.A)$$

As a tachyon at rest travels along $x_1$ but not along $x_0$, $v \to \infty$ for a tachyon and we obtain

$$X_0'^{\sim} = -iX_1, \qquad X_1'^{\sim} = -iX_0 \qquad (4.1.B)$$

We call this *the tachyonic transformation*.

Next we prove that the versatile Dirac and photon equations are invariant under this transformation for four-vector behaviour of the bispinors $\phi_1$ and $\phi_2^\ddagger$. Let

$$\mathbf{X}^{\sim} = \mathbf{i}_0 X_0^{\sim} + \mathbf{i}_1 X_1 + \mathbf{i}_2 X_2 + \mathbf{i}_3 X_3, \quad (4.1.\text{C})$$
$$\mathbf{Y}^{\sim\ddagger} = \mathbf{i}_0 Y_0^{\sim} - \mathbf{i}_1 Y_1 - \mathbf{i}_2 Y_2 - \mathbf{i}_3 Y_3$$

Let $\mathbf{X}^{\sim}$ and $\mathbf{Y}^{\sim}$ be associated with two four-vectors, $\mathbf{X} = (iX_0^{\sim}, X_1, X_2, X_3)$ and $\mathbf{Y} = (iY_0^{\sim}, Y_1, Y_2, Y_3)$, with $X_0^{\sim}$ and $Y_0^{\sim}$ imaginary. Applying the tachyonic transformation, equation (4.1.B), to $\mathbf{X}$ and $\mathbf{Y}$ transforms the quaternions $\mathbf{X}^{\sim}$ and $\mathbf{Y}^{\sim\ddagger}$ according to

$$\mathbf{X}'^{\sim} = -\mathbf{i}_0 X_1 + \mathbf{i}_1 X_0^{\sim} + \mathbf{i}_2 X_2 + \mathbf{i}_3 X_3, \quad (4.1.\text{D})$$
$$\mathbf{Y}'^{\sim\ddagger} = -\mathbf{i}_0 Y_1 - \mathbf{i}_1 Y_0^{\sim} - \mathbf{i}_2 Y_2 - \mathbf{i}_3 Y_3$$

We have already introduced the treatment of a Lorentz transformation as a rotation in a spacetime with the temporal co-ordinate imaginary[1] and we treat the tachyonic transformation as a similar rotation here. We find that

$$\mathbf{X}'^{\sim} = \left(\frac{1+\mathbf{i}_1}{\sqrt{2}}\right)\mathbf{X}^{\sim}\left(\frac{1+\mathbf{i}_1}{\sqrt{2}}\right), \quad (4.1.\text{E})$$
$$\mathbf{Y}'^{\sim\ddagger} = \left(\frac{1-\mathbf{i}_1}{\sqrt{2}}\right)\mathbf{Y}^{\sim\ddagger}\left(\frac{1-\mathbf{i}_1}{\sqrt{2}}\right)$$

as may be easily checked. If we then define

$$\underline{\mathbf{X}}^{\sim} = \underline{\mathbf{X}}^{\sim}(\mathbf{X}^{\sim}, \mathbf{Y}^{\sim\ddagger}), \quad \mathbf{R}_T = \frac{1+\mathbf{i}_1}{\sqrt{2}}, \quad \mathbf{R}_T | (\mathbf{R}_T, \mathbf{R}_T^{\ddagger}) \quad (4.1.\text{F})$$

we may write

$$\underline{\mathbf{X}}'^{\sim} = \mathbf{R}_T | \underline{\mathbf{X}}^{\sim} \mathbf{R}_T |^{\ddagger} \quad (4.1.\text{G})$$

where $\underline{\mathbf{X}}'^{\sim}$ is the tachyonic transform of $\underline{\mathbf{X}}^{\sim}$.

We may transform the versatile Dirac and photon equation to the dashed frame by applying equation (4.1.G) to all the reflectors present. It is clear from the form of equation (4.1.G) that both the Dirac and photon equations are invariant when the variables undergo such a transformation. The form of equation (4.1.G) also tells us that we may use it for general tachyonic transformations in which $\mathbf{R}_T$ may take the value of any real quaternion of unit modulus and not only the ones which correspond to a relative velocity along $x_1$. There is no significant difference between equation (4.1.G) and the equation we used previously[1] to provide Lorentz transformations for finite velocities and further details may be found in the cited paper.

## 4.2 Applying the symmetry

We shall discuss the world lines of several particles, each of which corresponds to some straight or circular path in *L*. We develop a set of equations that enable us to calculate the influence of the electromagnetic fields on the trajectories easily.

From equations (3.2.D) and (3.2.E) for the first of equations (4.2.A); the third of equations (3.2.A) and equation (3.2.C) for the second; equations (3.2.D and 3.2.F) for the third; equation (2.3.O) and the definition of $v^{b\sim}$ for the fourth,

$$v^{y\sim} = \frac{\mu^y}{\eta^{y\sim}}, \qquad (m^{y\sim})^2 = (\mu^y)^2 + (\eta^{y\sim})^2 \qquad (4.2.A)$$

$$\eta^{y\sim} v^{y\sim} = (m^{y\sim})^2, \qquad v^{y\sim} = \frac{\delta s_1^y}{\delta s_0^{y\sim}}$$

where we have set the superscript *b* in the original equations to *y* because we also use these equations for other particles besides the Bohr electron. If a particle obeys equations (4.2.A) in a space of type *M* or *S*, its behaviour in *L* is described by the Dirac equation with an inverse-distance potential, since a constant potential in *M* or *S* can be deduced. However, the length of the circular orbit cannot be determined without some condition like that given by Bohr's second equation to fix the velocity.

From equations (4.2.A) we obtain

$$v^{\tilde{y}} \delta s_0^{\tilde{y}} = \eta^{\tilde{y}} \delta s_0^{\tilde{y}} + \mu^{y} \delta s_1^{y} \tag{4.2.B}$$

where $iv^{\tilde{y}}$ is the energy of the interaction and $is_0^{\tilde{y}}$ the matching temporal co-ordinate. It does not matter in what frame we calculate the right-hand-side since it represents the dot product of two four-vectors but $v^{\tilde{y}}$ and $s_0^{\tilde{y}}$ refer to the rest frame of the interaction.

We shall let dashed co-ordinates continue to indicate the tachyonic frame throughout. In particular $(s_0^{\tilde{y}}, s_1^{\tilde{y}})$ becomes $(s_0'^{\tilde{y}}, s_1'^{\tilde{y}})$ in this frame. We define

$$s_0'^{y} = is_0'^{\tilde{y}}, \qquad s_1'^{y} = is_1'^{\tilde{y}} \tag{4.2.C}$$

From equations (4.1.B), we obtain

$$s_0'^{\tilde{y}} = -is_1^{y}, \qquad s_1'^{\tilde{y}} = -is_0^{y} \tag{4.2.D}$$

Using the first of equations (4.2.C) for the first, and the first of equations (2.3.O) for the second, we obtain

*The two-body interaction with a circle in time*

$$s_0'^y = s_1^y, \qquad s_1'^{y\sim} = s_0^{y\sim} \qquad (4.2.\text{E})$$

In the dashed frame an energy-momentum four-vector, $(\eta^y, \mu^y)$, becomes $(\eta'^{y\sim}, \mu'^{y\sim})$. We define

$$\eta^y = i\eta^{y\sim}, \quad \mu^y = i\mu^{y\sim}, \quad \eta'^y = i\eta'^{y\sim}, \quad \mu'^y = i\mu'^{y\sim} \qquad (4.2.\text{F})$$

Using the tachyonic transformation, equations (4.1.B), we obtain

$$\eta'^y = \mu^y, \qquad \mu'^{y\sim} = \eta^{y\sim} \qquad (4.2.\text{G})$$

We may solve the tachyonic form of the Dirac equation and arrive at the tachyonic form of equations (4.2.A). Equation (4.2.B) then becomes

$$v'^y \delta s_0'^y = \eta'^y \delta s_0'^y + \mu'^{y\sim} \delta s_1'^{y\sim} \qquad (4.2.\text{H})$$

$-iv'^y$ is the energy of the interaction in the dashed rest frame of the interaction, from the third of equations (4.2.F) with $v$ in place of $\eta$. $s_0'^{y\sim}$ is the matching temporal co-ordinate from the first of equations (4.2.C). Since the right-hand side of equation (4.2.H) represents the dot product of two four-vectors,

$$\eta'^y \delta s_0'^y + \mu'^{y\sim} \delta s_1'^{y\sim} = \eta^{y\sim} \delta s_0^{y\sim} + \mu^y \delta s_1^y \qquad (4.2.\text{I})$$

From equations (4.2.B) and (4.2.H)

$$v'^y \delta s_1^y = v^{y\sim} \delta s_0^{y\sim} \qquad (4.2.\text{J})$$

## 5. COUPLED INTERACTION

### 5.1 Angular momentum

We want to find the angular momentum and energy of the coupled Bohr and circle interactions for an observer in the rest frame of the nucleus in $L$. Since the circle interaction has no angular momentum in the undashed frame, the total is supplied by the Bohr interaction and we do not need to consider the circle interaction at all here. The Bohr electron has the same orbital angular momentum in $L$ using the momentum in $S^b$, as it did using the momentum in $M$ for the calculation we described previously.[2] This follows because the previous solution for the Bohr electron using $M$ and the solution in $S^b$ given in section 3.2 are identical. Further details may be found in the cited paper.

### 5.2 Coupled interaction in $S^h$

We demonstrate that we may donate the energy of the circle and Bohr interactions to a single particle, which obeys the Dirac equation in a space of type $S$ we shall call $S^h$. We call the particle *the heavy-electron*. We start with the Bohr electron. Equation (4.2.B) with $y = b$ gives

$$v^{b\sim}\delta s_0^{b\sim} = \eta^{b\sim}\delta s_0^{b\sim} + \mu^b \delta s_1^b \qquad (5.2.A)$$

We add in the contribution of the circle interaction, setting

$$v^{h\sim} = v^{b\sim} + \eta^{l\sim} \qquad (5.2.B)$$

and obtain from equation (5.2.A)

$$v^{h\sim}\delta s_0^{b\sim} = \left(\eta^{b\sim} + \eta^{l\sim}\right)\delta s_0^{b\sim} + \mu^b \delta s_1^b \qquad (5.2.C)$$

However, although equation (5.2.C) is of the same form as equation (4.2.B), it does not represent a solution of the Dirac equation with an inverse-distance potential in *L*, because the first three of equations (4.2.A) are not simultaneously obeyed. We can transform (5.2.C) into a form that does represent a solution for the heavy electron with an increased mass of $im^{h\sim}$

$$v^{h\sim}\delta s_0^{b\sim} = \eta^{h\sim}\delta s_0^{b\sim} + \mu^h \delta s_1^b \qquad (5.2.D)$$

where $i\eta^{h\sim}$ is the energy and $\mu^h$ the momentum of the heavy electron, $iv^{h\sim}$ is the total energy associated with the interaction of the heavy electron and the nucleus and

$$\eta^{h\sim} = \frac{v^{h\sim}(\delta s_0^{b\sim})^2}{(\delta s_0^{b\sim})^2 + (\delta s_1^b)^2}, \qquad \mu^h = \frac{v^{h\sim}\delta s_1^b \delta s_0^{b\sim}}{(\delta s_0^{b\sim})^2 + (\delta s_1^b)^2}, \qquad (5.2.E)$$

$$m^{h\sim} = \frac{v^{h\sim}\delta s_0^{b\sim}}{\sqrt{(\delta s_0^{b\sim})^2 + (\delta s_1^b)^2}}$$

Equations (4.2.A) are now obeyed with $y = b$ for $s_0^{y\sim}$ and $s_1^y$ and $y = h$ otherwise. At this point we find the origin of the energy required for the circle wave in the potential appearing in the Dirac equation whose solution is described by equation (5.2.D).

We quantised the angular momentum of the coupled interaction by insisting on equation (3.3.B) for the Bohr electron. We have the freedom to do this for only one space. We expect that the addition of a temporal circle should change only the energy and not the angular momentum. This means that we must quantise the angular momentum of the Bohr rather than the heavy electron.

We shall want the total energy in the dashed frame. We may convert to this frame using equation (4.2.J) with the same conditions on $y$, except that we will set $y = m$ for the superscript on $v'$ for reasons that become apparent later. We obtain

$$v'^m \delta s_1^b = \eta^{h\sim} \delta s_0^{b\sim} + \mu^h \delta s_1^b \quad (5.2.F)$$

where $-iv'^m$ is the energy of the total interaction taking $-is_1^b$ as the temporal co-ordinate. We may now convert the right-hand-side back into the same form as the right-hand-side of equation (5.2.C) since the two are equal

$$v'^m \delta s_1^b = \left(\eta^{b\sim} + \eta^{l\sim}\right)\delta s_0^{b\sim} + \mu^b \delta s_1^b \quad (5.2.G)$$

We find $v'^m$ by applying the first, second and fourth of equations (4.2.A) with $y = b$, obtaining

$$v'^m = \frac{\left(m^{b\sim}\right)^2}{\mu^b} + \frac{\eta^{l\sim}}{v^{b\sim}} \quad (5.2.H)$$

## 5.3 Coupled interaction in $M^m$ and $L$

We need to find the energy in space of type $M$, which we call $M^m$ here, in parallel with the previous calculation of energy made for the Bohr interaction.[2] As previously, the energy operator operates on the temporal co-ordinate, $x_0^m$, of an electron obeying the Dirac equation in $M^m$ for an inverse-distance potential in $L$. This is the coupling electron. $L$ and $M^m$ share the temporal plane in the undashed rest frame of the nucleus and therefore $x_0^m$, as previously, while $M^m$ and $S^h$ share the spatial plane in the same frame. We identify the heavy electron with the coupling electron in the spatial plane. The associated energies may differ because the two do not share a temporal plane.

Let the first two co-ordinates of the coupling electron in $M^m$ be $(x_0^m, s_1^b)$ in the rest frame of the nucleus. Since the Dirac equation describes the coupling electron in $M^m$, equations (4.2.A) apply. For these we have $y = m$ except for the mass, $m^y$, which we set to $m^b$, $s_0^y$ and $s_0^{y\sim}$ which we set to $x_0^m$ and $x_0^{m\sim}$ with $ix_0^{m\sim} = x_0^m$, and $s_1^y$ which we set to $s_1^b$. We call these *the combination conditions*. From equations (4.2.A) with the combination conditions we obtain the analogue of equation (4.2.B)

$$v^{m\sim}\delta x_0^{m\sim} = \eta^{m\sim}\delta x_0^{m\sim} + \mu^m \delta s_1^b \qquad (5.3.\text{A})$$

where $iv^{m\sim}$ is the energy of the coupled interaction between the coupling electron and the nucleus if we take $x_0^m$ as the temporal co-ordinate. $i\eta^{m\sim}$ and $\mu^m$ are the energy and momentum of the coupling electron in $M^m$.

We cannot quantise the angular momentum at this point either, because if we did we would reproduce the Bohr interaction without the extra energy added by the circle interaction.

Converting equation (5.3.A) to the dashed frame and using equation (4.2.J) with the combination conditions, we obtain

$$v'^m \delta s_1^b = \eta^{m\sim}\delta x_0^{m\sim} + \mu^m \delta s_1^b \qquad (5.3.\text{B})$$

where $v'^m$ is common to the interactions of both the heavy and coupling electron since $s_1^b$ is. This follows because $s_1^b$ would appear in the operator used to determine $v'^m$. Applying the first, second and fourth of equations (4.2.A) with the combination conditions leads to

$$v'^m = \frac{(m^{b\sim})^2}{\mu^m} \qquad (5.3.C)$$

From equation (5.2.H) we obtain

$$\frac{(m^{b\sim})^2}{\mu^m} = \frac{(m^{b\sim})^2}{\mu^b} + \frac{\eta^{l\sim}}{v^{b\sim}} \qquad (5.3.D)$$

Substituting from equation (3.2.E) for $\mu^b$, from equation (3.1.D) for $\eta^{l\sim}$, calculating $\mu^m$ in terms of $-iv^{m\sim}$, the velocity of the coupling electron, from the first and second of equations (4.2.A) with the combination conditions and simplifying, we obtain

$$\frac{\sqrt{1+(v^{m\sim})^2}}{v^{m\sim}} = \frac{\sqrt{1+(v^{b\sim})^2}}{v^{b\sim}} + \frac{n_r}{n_\theta v^{b\sim}} \qquad (5.3.E)$$

a purely geometrical relationship.

From the first, second and third of equations (4.2.A) with the combination conditions we obtain an expression for the total energy of the coupled interaction, $iv^{m\sim}$,

$$v^{m\sim} = m^{b\sim}\sqrt{1+(v^{m\sim})^2} \qquad (5.3.F)$$

This is similar in form to the expression for the total energy of the Bohr interaction, $iv^{b\sim}$, given in equation (3.2.F). We solve to find the energy of the coupled interaction, $iv^{m\sim}$, using equation (5.3.E),

$$v^{m\sim} = m^{b\sim}\left\{1 - \frac{1}{\left(\frac{\sqrt{1+(v^{b\sim})^2}}{v^{b\sim}} + \frac{n_r}{n_\theta v^{b\sim}}\right)^2}\right\}^{-\frac{1}{2}} \quad (5.3.G)$$

Substituting for $v^{b\sim}$ from equation (3.3.C), we obtain the final formula for the coupled interaction energy levels, $iv^{m\sim}$,

$$v^{m\sim} = m^{b\sim}\left\{1 + \frac{\alpha^2}{\left(\sqrt{n_\theta^2 - \alpha^2} + n_r\right)^2}\right\}^{-\frac{1}{2}} \quad (5.3.H)$$

These energy levels are the same as those predicted by both Sommerfeld's model of the one-electron atom [8, 11] and the model using the Dirac interaction. [5, 8] The same is true for the angular momentum eigenvalues we discussed in section 5.1. In particular, the full spectrum of energy levels including the fine structure appears. The correspondence between $n_\theta$ and $n_r$ and the quantum numbers labelling the Dirac interaction energy and angular momentum eigenvalues is the same as for Sommerfeld's model. [8]

**5.4   Revisiting the Bohr interaction**

Equation (5.3.H) reduces to give equation (3.2.F) with $v^{b\sim}$ in place of $v^{m\sim}$ when $n_r$ is zero. We can therefore apply our method in sections 5.2 and 5.3 to the Bohr interaction alone when the circle is not vibrating. This means that we can suppose that when the two-body interaction is in its ground state or excited to one of the principle levels the structure also embodies a temporal as well as a spatial circle. In this case the space $M^m$ becomes the space $M$ that we have already explored extensively.[2, 3]

## 6. DERIVATION OF QED

In a previous paper[3] we derived QED from the two-body interaction, on the assumption that the charge density current and potential field were sufficiently well behaved. In this section we refer to this previous paper where not otherwise indicated. The derivation could be done using either the ground state or an excited state for the two-body interaction but excited states were restricted to the principal energy levels. Here, we sketch the inclusion of fine structure by adding the circle interaction. The principle of the method remains unaltered.

The quasi-particle associated with the circle wave is neutral and so the net charge introduced by the circle interaction is zero. This means that we can extend the existing results trivially for the photon equation.

For the Dirac equation, the starting point of the existing method is to show the equation applies, if the Bohr equations do for the Bohr interaction, by explicit construction of a solution for the spaces $M_q$. There is one such space, of type $M$, for every point, $q$, in $L$. We follow a similar course here but for the coupled rather than the Bohr interaction. We use the spaces $M_q^m$, with one space, $M^m$, for every point, $q$, in $L$, in a similar notation to our previous one. The coupling electron satisfies a Dirac equation whose solution is described by equation (5.3.A). It is of similar form to equation (2.3.M) but for the space $M_q^m$. Following the guidelines set out in section 5, we can calculate the value of the variables required to assemble the wave function at $q$, $\underline{\Phi}^B = \underline{\Phi}^{m_q}\left(\phi_1^{m_q}, \phi_2^{m_q}\right)$,

$$\phi_1^{m_q} = \exp\left\{i\left(v^{m_q\sim}\tilde{x}_0^q + \underline{\mu}^m\ s_1^{b_q}\right)\right\} \quad (6.A)$$

$$\phi_2^{m_q} = i\left(v^{m_q\sim} - \mathbf{i}_{s_1} - eA^{m_q\sim}\right)\left(\mathbf{M}^\sim\right)^{-1}$$
$$\exp\left\{i\left(v^{m_q\sim} x_0^{m_q\sim} + \underline{\mu}^m\ s_1^{b_q}\right)\right\}$$

where $iA^{m_q\sim}$ is the potential. We have added a subscript, $q$, to indicate that in general the variables have different values at different points, $q$, in $L$; otherwise the variables are the same as those discussed in section 5. Equation 6.A implies Bohr's first equation,

$$\frac{ie^2}{R_1^{m_q}} = \frac{m^{b\sim}\left(v^{m_q\sim}\right)^2}{\sqrt{1+\left(v^{m_q\sim}\right)^2}} \quad (6.B)$$

where $2\pi R^{m_q}$ is the length of the circular orbit. Since we can calculate the velocity, $-iv^{m_q\sim}$, from equation (5.3.E), $R^{m_q}$ is known. The conservation of energy argument we discussed previously[2, 3] means that the coupling electron can appear in $L$ at a distance $R^{m_q}$ from the nucleus on a circle or its sphere of revolution.[3] We note here that this means the spatial and temporal symmetries are the same as those of the Bohr interaction,[2] since we may ignore the circle interaction for angular momentum eigenstates.

We also derived[3] an equation connecting the variables at point $q$ for the Bohr interaction and showed that the equation was physically reasonable. It described the interaction of the Bohr electron and a quasi-particle. The quasi-particle arose from the charge density, $\rho$, which obeyed the photon equation and resulted in a potential, $A$. The electron obeyed the Dirac equation with this potential. The equation relating the variables at $q$ was

$$\frac{\rho^2}{de^2} - A^3\rho - (m^b)^2 dA^4 = 0 \tag{6.C}$$

where

$$d = \frac{3\pi}{n^2 h^2} \tag{6.D}$$

$n$ is the equivalent of the $n_\theta$ discussed here and $n_r$ was zero. We can see, by comparing equation (5.3.H) with equation (3.2.F) with the value of $v^{b\sim}$ given by (3.3.C) inserted, that in this paper we have made the replacement

$$\sqrt{n_\theta^2 - \alpha^2} \to \sqrt{n_\theta^2 - \alpha^2} + n_r \tag{6.E}$$

in going from the Bohr electron and interaction to the coupling electron and coupled interaction, leading to a new value for $d$, $d'$, where

$$d' = \frac{3\pi}{h^2\left(n_\theta^2 + n_r^2 + 2n_r\sqrt{n_\theta^2 - \alpha^2}\right)} \tag{6.F}$$

which is always positive if $n_r$ is. Equation (6.C) may be solved for $\rho$ with $d'$ instead of $d$ giving

$$\rho = \frac{A^2 e^2 d'}{2}\left(A \pm \sqrt{A^2 + \frac{4(m^b)^2}{e^2}}\right) \tag{6.G}$$

The discussion of physically reasonable values for $\rho$ and $A$ then proceeds for non-zero $n_r$ as it did before for zero $n_r$, provided we assume

$$n_\theta \sim 1, \qquad n_r \sim 1 \qquad (6.\text{H})$$

where ~ stands for *is commensurate with*. The rest of the previous discussion in which we showed that the photon and Dirac equations are valid for all points in $L$ also follows on, with $M_q$ replaced by $M_q^m$.

We showed previously that the quantum theory is QED in the instance where the two-body interactions in the $M_q$ are given by the Bohr equations. Here we have generalised this by showing that the quantum theory is also QED if the energy given by the Bohr equations is augmented by the energy due to the vibration of a temporal circle obeying equation (3.1.D). Now we have an inverse, that the interaction of a potential field obeying the photon equation and a particle obeying the Dirac equation may always be pictured instead as a sea of infinitesimal two-body interactions, the coupled interactions on the $M_q^m$. This is true because the solution to the Dirac equation describing the behaviour of the first-quantised bound state is never more general than the one that we explored in section 5, as the Dirac interaction demonstrates.

## 7. DISCUSSION

In this paper we completed our previous discussion[2] of new spaces, of type $M$, $T$ and $S$, on which the photon and Dirac equations hold. The spaces provide a set of new solutions for the equations by altering the metric and topology that applies to our Minkowski spacetime, $L$. We have examined one of these new solutions, the coupled interaction that corresponds to the inverse-distance-squared law that we know holds in *L* for the strength of the electric field of a point charge. The other solutions correspond to different physical laws, which it is more difficult to realise experimentally. The coupled interaction can be interpreted to yield the familiar eigenvalues, including fine structure, of the usual model of the two-body interaction, the Dirac interaction.[5]

We used first-quantised solutions here. Although the second-quantised solution is already well known,[10] it would still be of interest to find a similar series using the coupled interaction directly. This is because it has new symmetries, for example, cylindrical symmetry for the angular momentum eigenstate,[2] that might be respected in such a full solution.

We also completed our previous discussion[3] of a new method of quantising the classical electromagnetic field by quantising the two-body interaction. We have applied this to quantising General Relativity.[4]

## ACKNOWLEDGEMENTS
One of us (Bell) would like to acknowledge the assistance of E. A. E. Bell.